\documentclass{pnastwo}
\usepackage{bm}
\usepackage{graphicx}
\usepackage{amsmath}
\usepackage{amssymb}
\usepackage{color}

\begin{document}
\title{Magnetism and superconductivity driven by identical 4$f$
states in a heavy-fermion metal}
\author{Sunil Nair\affil{1}{Max-Planck-Institut f\"ur Chemische
Physik fester Stoffe, D-01187 Dresden, Germany}, O.
Stockert\affil{1}{}, U. Witte\affil{2}{Institut f\"ur
Festk\"orperphysik, TU Dresden, D-01062 Dresden,
Germany}\affil{3}{Helmholtz-Zentrum Berlin f\"ur Materialien und
Energie, D-14109 Berlin, Germany}, M. Nicklas\affil{1}{}, R.
Schedler\affil{3}{}, K. Kiefer\affil{3}{}, J. D.
Thompson\affil{4}{Los Alamos National Laboratory, Los Alamos, NM
87545, USA}, A. D. Bianchi\affil{5}{University of California,
Irvine, CA 92697, USA}, Z. Fisk\affil{5}{}, S. Wirth\affil{1}{},
\and F. Steglich\affil{1}{}} \maketitle

\begin{article}
\begin{abstract}
The apparently inimical relationship between magnetism and
superconductivity has come under increasing scrutiny in a wide
range of material classes, where the free energy landscape
conspires to bring them in close proximity to each other. This is
particularly the case when these phases microscopically
interpenetrate, though the manner in which this can be
accomplished remains to be fully comprehended. Here, we present
combined measurements of elastic neutron scattering,
magnetotransport, and heat capacity on a prototypical heavy
fermion system, in which antiferromagnetism and superconductivity
are observed. Monitoring the response of these states to the
presence of the other, as well as to external thermal and magnetic
perturbations, points to the possibility that they emerge from
different parts of the Fermi surface. This enables a single 4$f$
state to be both localized and itinerant, thus accounting for the
coexistence of magnetism and superconductivity.
\end{abstract}

\dropcap{T}he ground state properties of a system are of
fundamental importance and the starting point for considering the
excitations that enliven real systems. The prevalent electronic
ground states of metals, magnetism and superconductivity, are
typically mutually exclusive quantum many body phenomena. This
antagonism can be evaded by spatial separation (e.g. in some
Chevrel phases \cite{fis}) or by subdividing the 5$f$ states in
some actinide compounds into more localized and more itinerant
parts giving rise to magnetism or participating in
superconductiviy, respectively (see, e.g. \cite{sato}). The quest
for microscopic coexistence of these phenomena involving {\em
identical} electrons is fueled by the expectation for insight into
the complex behavior of new materials with intertwined ground
states as, e.g., the cuprate superconductors. Experimentally, this
not only requires finding an appropriate material, but also calls
for a concerted investigation of both the charge and the spin
channel and hence, judiciously chosen measurement methods.

The heavy fermion metals offer an interesting playground where
magnetism and superconductivity can both compete and coexist. In
these systems, the hybridized \textit{f} electrons are not only
responsible for long-range magnetic order, but are also involved
in superconductivity. In this context the Ce$M$In${_5}$ (where $M$
= Co, Ir or Rh) family of heavy fermion metals has been in vogue
due to their rich electronic phase diagrams in which an intricate
interplay between superconductivity and magnetism is observed
\cite{joe}. For instance, in CeCoIn$_5$, a superconducting ground
state is found below a transition temperature $T_{\rm c} \approx$
2.3 K whereas CeRhIn${_5}$ orders antiferromagnetically below
$T_{\rm N} \approx 3.7$~K \cite{joe}. On the other hand,
superconductivity is observed in the latter compound by
application of pressure whereas the proximity to magnetism in
CeCoIn$_5$ is demonstrated by the likely existence of a zero
temperature magnetic instability \cite{joe}. Moreover, neutron
scattering experiments indicate strong antiferromagnetic (AF)
quasielastic excitations at wavevectors $Q =
(\frac{1}{2}~\frac{1}{2}~\frac{1}{2})$ and equivalent positions in
the paramagnetic regime \cite{broholm}. However, the excitations
become fully inelastic when entering the superconducting state,
resulting in the appearance of a spin resonance. These findings
underline the analogy to the cuprate high-temperature
superconductors \cite{super,sup1}.

We conducted a comprehensive investigation of the magnetic order
and superconductivity in CeCo(In$_{0.9925}$Cd$_{0.0075}$)$_5$.
Neutron scattering, magnetotransport and heat capacity, i.e.,
microscopic and macroscopic, spin and charge sensitive studies
have been combined on flux-grown single crystals of the same
batch. These combined efforts not only allow to unambiguously pin
down the associated effects but also to cross-fertilize the
methods. We find a local duality of the electronic 4$f$ degrees of
freedom implying electronic phase separation on the Fermi surface.

\begin{figure}[b]
\centering \includegraphics[width=8.5cm,clip]{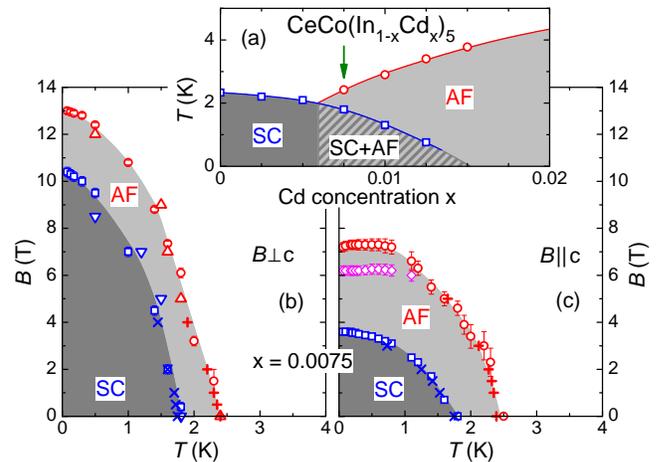}
\caption{\label{fig1}(a) Doping $x$ dependence of
antiferromagnetic (AF) and superconducting (SC) transition
temperatures in CeCo(In$_{1-x}$Cd$_x$)$_5$ for Cd-content $x \le
0.02$ (after Ref.~\cite{pham}). The crystals investigated
here (arrow) exhibit both transitions. (b) $B$--$T$ phase diagram
of CeCo(In$_{0.9925}$Cd$_{0.0075}$)$_5$ obtained from
magnetotransport ($\circ,\square$), neutron scattering
($\triangle, \bigtriangledown$) and heat capacity (+,$\times$)
measurements with $B\! \perp\! c$. (c) $B$--$T$ diagram for $B\!
\parallel\! c$ from magnetotransport and heat capacity.
Indications of a transition within the AF phase are found
($\diamond$).}
\end{figure}

The specific composition with $x =$ 0.0075 was chosen since
$T_{\rm c} \approx 1.7$ K and $T_{\rm N}\approx 2.4$~K are closest
within the CeCo(In$_{1-x}$Cd$_x$)$_5$ series \cite{pham},
Fig.~\ref{fig1}(a). Consequently, the interplay between
superconductivity and antiferromagnetism is expected to be most
pronounced \cite{kato}. Earlier studies \cite{curro,nicklas} were
conducted on samples with $x \geq 0.01$ where $T_{\rm c}$ is small
compared to $T_{\rm N}$. Thus, the condensation of conduction
electrons into Cooper pairs effectively took place in a state
where fluctuations of the AF order parameter were not appreciably
large, i.e., where the balance of the two phenomena is already
shifted toward AF order. As a result, no significant anomaly in
the magnetic intensity as determined by neutron scattering was
observed on entering the superconducting regime \cite{nicklas}.

The resulting magnetic field--temperature ($B$--$T$) phase
diagrams are presented in Fig.~\ref{fig1}(b) and (c) for $B\!
\perp\! c$ and $B \! \parallel \! c$, respectively. The excellent
agreement of results obtained by three very different methods
evidences that bulk properties are probed. The strikingly
equivalent behavior of the superconducting and AF phase boundary,
in particular for $B\! \perp\! c$, is indicative of a mutual
influence of the two phenomena. The steep initial slope of $T_{\rm
c}(B)$ of approximately $-13(-4)$ T/K for $B\!
\perp\!\!(\parallel)\, c$ indicates a large effective
quasiparticle mass, i.e., heavy fermion superconductivity.

\begin{figure}[t]
\centering \includegraphics[width=8.4cm,clip]{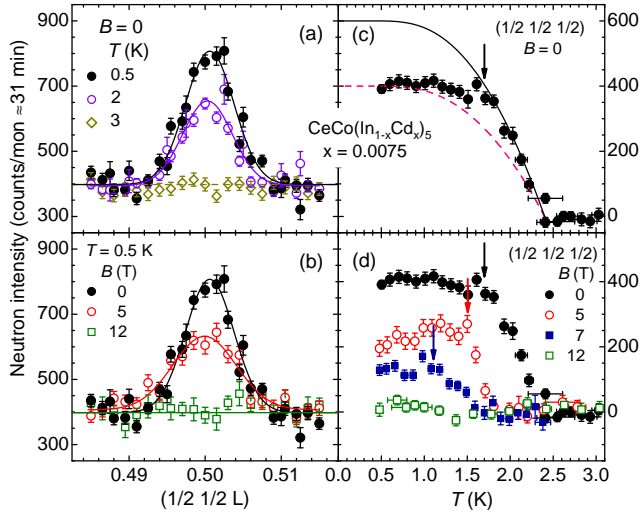}
\caption{\label{fig2} Elastic neutron scattering scans in
CeCo(In$_{0.9925}$Cd$_{0.0075}$)$_5$ along $[001]$ and across
$(\frac{1}{2}~\frac{1}{2}~\frac{1}{2})$: (a) at different
temperatures in zero magnetic field and (b) for several magnetic
fields at $T = 0.5$ K. Solid lines indicate fits with Gaussian
lineshape to the data. (c) Temperature dependence of the magnetic
intensity at $(\frac{1}{2}~\frac{1}{2}~\frac{1}{2})$ in $B=0$
along with fits from mean-field expectations to the data (solid
and dashed lines, see text). (d) Same as (c) in addition with data
for several magnetic fields. The paramagnetic background
contribution was subtracted from data in (c) and (d).}
\end{figure}

Initial elastic scans at the lowest temperature ($T=0.5$ K) across
the nuclear peaks confirmed the tetragonal crystal structure, with
lattice parameters $a = 4.595$\,{\AA} and $c = 7.533$\,{\AA}. To
search for magnetic intensity, scans along high-symmetry
directions were performed. Well below $T_{\rm N} \approx 2.4$ K
and $T_c \approx 1.7$ K, scans along $(\frac{1}{2}~\frac{1}{2}~l)$
revealed additional magnetic intensity at
$(\frac{1}{2}~\frac{1}{2}~\frac{1}{2})$ and symmetry equivalent
positions, see data at $T=0.5$ K in Fig.\,\ref{fig2}(a). Above
$T_{\rm N}$ this magnetic superstructure peak completely vanished,
{\it cf.} Fig.\,\ref{fig2}(a) for data at $T = 3$ K. The scans did
not indicate additional intensity at other, e.g. incommensurate,
positions. In particular, no magnetic superstructure peaks were
detected around $(\frac{1}{2}~\frac{1}{2}~0.3)$ or
$(\frac{1}{2}~\frac{1}{2}~0.4)$, which have been observed in the
closely related system CeRhIn$_5$ \cite{bao}. The commensurate
magnetic structure is therefore in close agreement with that
reported earlier on the $1$\% Cd doped sample \cite{nicklas}.
Elastic scans across $(\frac{1}{2}~\frac{1}{2}~\frac{1}{2})$ at $T
= 0.5$ K and for several magnetic fields are displayed in
Fig.\,\ref{fig2}(b). Obviously, a magnetic field of $B = 12$ T
suffices to fully suppress antiferromagnetism at this temperature.
More importantly, the observation of a magnetic superstructure
peak in zero magnetic field well inside the superconducting state
clearly demonstrates the coexistence of AF order and
superconductivity on a microscopic scale. Based on our heat
capacity measurements we emphasize that both, the transition into
the AF ordered and the superconducting state, are bulk
transitions.

In order to scrutinize the possible influence of superconductivity
on the AF order, the magnetic intensity at
$(\frac{1}{2}~\frac{1}{2}~\frac{1}{2})$ was recorded as a function
of temperature for different magnetic fields, Figs.\,\ref{fig2}(c)
and (d). In zero magnetic field the magnetic intensity increases
below $T_{\rm N}$ and displays a kink at $T_{\rm c}$ (marked by
arrows) with no further change in intensity at lower temperatures.
For increasing magnetic field, $T_{\rm N}$ and the overall
magnetic intensity are reduced. No magnetic intensity was detected
for $B = 12$ T. The assignment of this kink to $T_{\rm c}$ is
corroborated by the magnetotransport and heat capacity
measurements. An attempt to fit the zero-field magnetic intensity
by a mean-field model for the sublattice magnetization (using a
Brillouin function for an effective spin-$\frac{1}{2}$ system)
fails to describe the whole temperature dependence, as indicated
by the dashed line in Fig.\,\ref{fig2}(c). On the other hand, a
fit restricted to the temperature range $T_{\rm c} < T < T_{\rm
N}$ reproduces these data reasonably well [solid line in
Fig.\,\ref{fig2}(c)] and results in an expected magnetic intensity
for $T \rightarrow 0$ of about $40$\% larger than the
experimentally observed saturation value. Obviously, the onset of
superconductivity {\em prevents a further rise} of magnetic
intensity below $T_{\rm c}$ {\em without suppressing} the AF order
itself.

\begin{figure}[t]
\centering \includegraphics[width=8.5cm,clip]{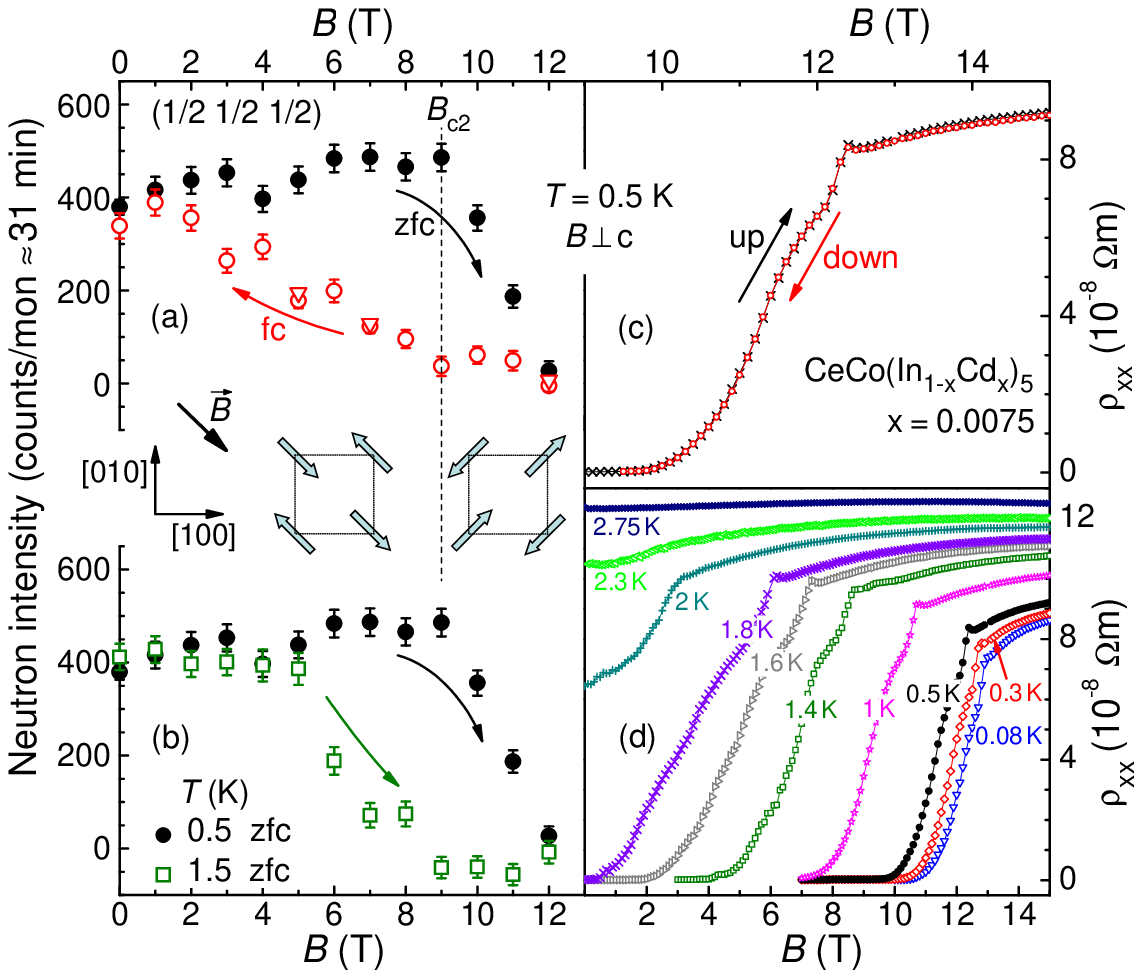}
\caption{\label{fig3} Comparison of neutron scattering and
resistivity measurements. (a) Field dependent magnetic intensity
at $(\frac{1}{2}~\frac{1}{2}~\frac{1}{2})$ and $T = 0.5$ K after
zero-field cooling (zfc) and field cooling (fc, $\triangledown$).
Arrows indicate the direction of magnetic field variation. The
different field conditions give rise to different domain
population as shown in the inset: For fc and $B > B_{\rm c2}$
($B_{\rm c2}$ is marked by the dashed line) only the depicted spin
configuration in the basal plane is found, below $B_{\rm c2}$ also
the second displayed domain (left) is increasingly occupied. (b)
Magnetic intensity after zfc at $T = 0.5$ K and 1.5 K. (c) Field
dependence of resistivity $\rho_{xx}$ for $B\! \perp\! c$. A
protocol analogous to (a) has been followed, yet no significant
hysteresis was found. (d) Resistivity $\rho{_{xx}}$ as function of
$B$ ($\perp\! c$) displayed for selected temperatures.}
\end{figure}

The magnetic intensities measured as a function of applied field
$B\parallel[1\overline{1}0]$ for different temperatures and
different protocols are directly compared to magnetotransport
$\rho_{xx}(B)$ in Fig.\,\ref{fig3} facilitating again a clear
assignment of the observed features. The disappearance of magnetic
intensity, signaling the transition from the antiferromagnetically
ordered phase to a paramagnetic one, nicely concurs with the
strong change in slope in $\rho_{xx}(B)$. On the other hand, the
kink in the field-dependent neutron intensity can be identified as
the superconducting upper critical field $B_{\rm c2}$ coinciding
with the approach to zero resistivity. The latter is also
supported by the similarity of the field-dependent neutron
intensity [Fig.~\ref{fig3}(b)] and its temperature dependence,
Fig.~\ref{fig2}(d).

Interestingly, a pronounced hysteresis is seen for the neutron
scattering intensities taken at increasing (zfc) and decreasing
magnetic field, Fig.~\ref{fig3}(a). Whereas the aforementioned
kink is observed for increasing magnetic field, in decreasing
field the magnetic intensity grows steadily and only reaches for
$B\! \rightarrow\! 0$ the values of the zero-field cooled
measurements. In the pristine CeCoIn$_5$, a multi-component ground
state (also discussed as a possible
Fulde-Ferrel-Larkin-Ovchinnikov phase) with characteristics of a
first order phase transition has been observed at low temperatures
($T<0.3$ K) in fields $B > 10$ T applied along the
$[1\overline{1}0]$ direction \cite{bianchi,science}. However, in
accord with the sensitivity of such a state to disorder its
existence in the Cd substituted system has been dismissed
\cite{yoshi}. It is to be noted that enhanced disorder arising
from Cd substitution increases the typical resistivity values in
this system by an order of magnitude in comparison to undoped
CeCoIn$_5$. Moreover, the range of magnetic fields within which
this hysteresis is observed in neutron scattering implies that the
hysteretic behavior is seen mainly above $B_{\rm c2}$ within the
AF phase, ruling out shielding effects. An alternative scenario
would involve that the field-driven transition from an AF phase
into a paramagnetic one is first order in nature. To investigate
this possibility, we have performed resistivity measurements in
slowly increasing and decreasing fields at $T = 0.5$ K. As shown
in Fig.~\ref{fig3}(c), no significant hysteresis is observed
indicating that the field-driven transition is continuous in
character (at least for $T \geq$ 0.06 K).

With the first order scenario likely ruled out, the observed
hysteresis in our neutron scattering data (and the lack of it in
$\rho{_{xx}}$) can only be explained by invoking the possibility
of two different domain populations in the field cooled and zero
field cooled measurements. Though relatively little explored in
comparison to ferromagnets, the existence of magnetic domains is
well established for anisotropic antiferromagnets. A particularly
well investigated example is elemental Cr for which the influence
of different domain populations as a function of measuring
protocols has been observed \cite{Cr1,Cr2}. Our neutron data
indicate an unequal domain population upon entering the
magnetically ordered state at low temperatures and high magnetic
fields. Decreasing the magnetic field at low $T$ and crossing the
phase boundary into the AF state, one domain configuration [with
magnetic moments $\perp \! B$, {\it cf.} Fig.\ \ref{fig3}(a)] is
strongly favored over the other (with magnetic moments $\parallel
\! B$) resulting in a substantially reduced magnetic intensity in
neutron scattering measurements (see {\em Materials and Methods}
section). Further reducing the magnetic field and inside the
superconducting state the second domain successively becomes
populated balancing the domain population when reaching $B = 0$,
identical to the zfc case. The associated domain walls strongly
influence the magnetotransport {\em only} if the electronic mean
free path $\ell$ is comparable to or larger than the domain wall
thickness $\delta$ \cite{Cr1}. The lack of hysteresis in our
transport measurements suggests that this criterion is not met in
the $B\! \perp\! c$ direction, i.e., $\delta \gg \ell$. Note that
even in undoped CeCoIn$_5$, $\ell$ is reduced to a few ten nm
already in moderate fields \cite{kas}. However, we were able to
resolve a tiny hysteresis ($\lesssim$ 0.2 T at 0.2 K) in the
magnetoresistance for $B\! \parallel\! c$ within the AF regime
\cite{sunil}. This observation is consistent with an enhanced
(factor of 1.5) dynamic spin correlation length within the $ab$
plane compared to the $c$ direction \cite{broholm,cur2} which
indicates a reduced $\delta$ in $c$ direction.

\begin{figure}[t]
\centering \includegraphics[width=8.2cm,clip]{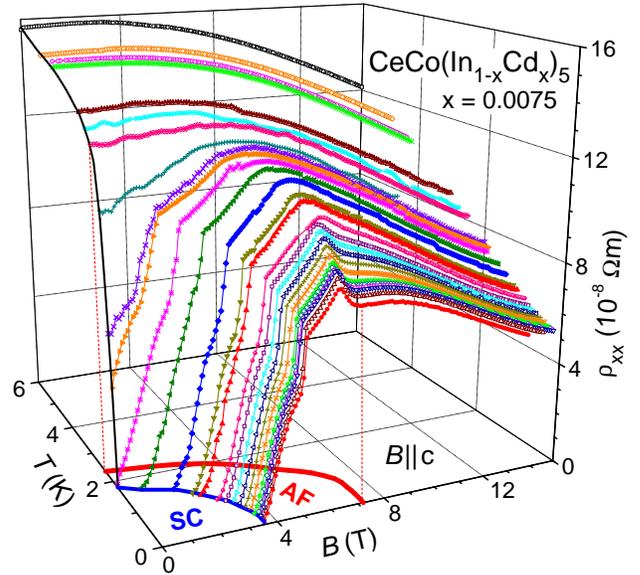}
\caption{\label{fig4} (a) Magnetic field dependence of the
resistivity $\rho{_{xx}}$ as measured in
CeCo(In$_{0.9925}$Cd$_{0.0075}$)$_5$ at different temperatures
with $B\! \parallel\! c$. The phase boundaries associated with
superconductivity and antiferromagnetism are marked.}
\end{figure}

In order to trace the anisotropic nature of the superconductivity
and magnetism in this system, we also measured $\rho{_{xx}}$ with
field applied along the crystallographic $c$ axis. This is shown
in Fig.\ \ref{fig4}, with the zero-resistance superconducting
state and the field-induced destabilization of AF order being
clearly marked out. At lowest temperature and for decreasing
field, the sharp increase in resistivity at $B \sim$~7 T indicates
carrier localization due to the onset of AF order. On further
reduction of the magnetic field, a drop in $\rho_{xx}(B)$ for $B
\lesssim$~6 T is observed within the AF state. This could possibly
originate from a spin rearrangement as found in CeCu${_5}$Au
\cite{paschke}, a scenario that would also account for the
observed anisotropy in $\rho_{xx}(B)$. Alternatively, one might
speculate that the drop in $\rho_{xx}(B)$ may be caused by a
change in ordering vector as, e.g., observed CeCu$_2$Si$_2$
\cite{geg}. The signatures of these two transitions merge as they
become broadened at increasing temperatures. The magnetoresistance
is negative down to $T =$ 0.06 K for 7 T $\lesssim B \leq$ 15 T
manifesting that there is {\em no} Fermi liquid regime in the
investigated field range. This effectively rules out the presence
of a quantum critical point in the $B\! \parallel\! c$ direction.
Interestingly, for $B\! \perp\! c$ the destruction of long-range
AF order is succeeded by a field range of positive
magnetoresistance which indicates that the system enters into a
regime with coherent Kondo scattering \cite{brandt}. Analyzing
$\rho_{xx}(T)$ for constant fields did not reveal any signature of
a $T^2$ dependence eliminating the possibility of Fermi liquid
behavior also for $B\! \perp\! c$ [a Kohler's scaling analysis is
hampered by large uncertainties in $\rho{_{xx}}$(0)].

The almost constant neutron intensity below $T_{\rm c}$ is
intriguing. Its analysis above and the electronic transport
measurements indicate a second order phase transition at $T_{\rm
c}$ without spatial phase separation. Then, the deviation of the
neutron intensity from its expected value below $T_{\rm c}$
implies coexistence and, more importantly, mutual influence of AF
and superconducting order. These conclusions go well beyond those
drawn from earlier Nuclear Magnetic Resonance measurements
\cite{curro}: Although the microscopic coexistence of AF and
superconducting order was inferred, the interplay between the two
different types of order was not observed. Based on our new
measurements we speculate that the low-energy magnetic
fluctuations are gapped by superconductivity and likely shifted to
higher energies (possibly to the resonance at 0.6 meV observed in
undoped CeCoIn$_5$ \cite{broholm}), a similar mechanism as
discussed for the cuprates \cite{dahm}. The delicate,
unprecedented balance of the two states may result from the
proximity of $T_{\rm c}$ and $T_{\rm N}$ in the chosen compound.
Since the commensurability of the AF order with $\tau =
(\frac{1}{2}~\frac{1}{2}~\frac{1}{2})$ and NMR studies
\cite{curro} suggest mainly local magnetism, the single 4$f$ state
spans both local and itinerant character in momentum space. We
note that the local coexistence of AF and superconducting order is
corroborated by the spin-spin correlation length $\xi_m$ clearly
exceeding the superconducting one, $\xi_{GL}$. The former can be
estimated from the broadened (beyond resolution) Gaussians of the
neutron scattering intensities ({\it cf.}, e.g.,
Fig.~\ref{fig2}(a)), $\xi_m \approx$ 32 nm, whereas an upper bound
$\xi_{GL} \lesssim$ 10 nm can be inferred from the estimated
critical field $B_{c2}(T\rightarrow 0)$.

In conclusion, magnetotransport, heat capacity and elastic neutron
scattering measurements were combined to unambiguously identify
the respective features of antiferromagnetic and superconducting
order in the heavy-fermion alloy CeCo(In$_{1-x}$Cd$_x$)$_5$ with
$x=0.0075$ resulting in a highly consistent $B$--$T$ phase
diagram. Below $T_{\rm c}$, superconductivity and magnetism
correlate via identical 4$f$ states resulting in a delicate
balance of local coexistence. The phase transitions appear to be
continuous in nature, and the pronounced hysteresis observed in
neutron scattering measurements likely arises from different
domain populations dictated by the sample history. The destruction
of antiferromagnetic order at lowest temperature is {\em not}
followed by Fermi liquid behavior for a substantial field range
leaving the ground state unresolved.\\

\begin{materials}
A 12 mg platelet-like single crystal was used for neutron
scattering as well as for heat capacity measurements in a
commercial Physical Property Measurement System with $^3$He
insert. Magnetoresistance measurements were conducted with $B \le$
15 T applied both $\parallel\! c$ and $\perp\! c$ for 0.06~K~$\leq
T \le$ 6 K.

Elastic neutron scattering measurements were conducted in the
temperature range 0.5 K $\leq T \leq 3$ K within the
$[110]\!-\![001]$ horizontal scattering plane, both in field
cooled (fc) and zero-field cooled (zfc) conditions. Magnetic
fields $B \leq$ 12 T were applied within the basal \textit{ab}
plane. These measurements were performed on the thermal
triple-axis spectrometer E1 at the BER-II reactor of the
Helmholtz-Zentrum Berlin f\"ur Materialien und Energie in
Berlin/Germany.

With respect to our discussion of magnetic domains it should be
noted that the neutrons can only couple to moments perpendicular
to the momentum transfer {\bf Q}. In our case, this allows to
unambiguously assign a largely reduced magnetic intensity at {\bf
Q} $= (\frac{1}{2}~\frac{1}{2}~\frac{1}{2})$ and high magnetic
fields $B$ to such domains with magnetic moments perpendicular to
$B$ since only for these domains there are components of the
magnetic moment parallel to {\bf Q}. In contrast, for the domains
with magnetic moments parallel to $B$ all individual moments are
aligned perpendicular to {\bf Q} which results in a much higher
neutron intensity.
\end{materials}

\begin{acknowledgments}
This work was partially supported by the DFG Research Unit 960
``Quantum Phase Transitions''. Z.F. acknowledges support through
NSF-DMR-071042. Work at Los Alamos National Laboratory was performed
under the auspices of the U.S. Department of Energy, Office of Science.
\end{acknowledgments}

\end{article}

\begin{thebibliography}{23}
\bibitem{fis} Fischer {\O} (1978) Chevrel phases - superconducting
and normal state properties. {\it Appl. Phys.} 16:1-28.

\bibitem{sato} Sato NK, {\it et al.} (2001) Strong coupling between
local moments and superconducting `heavy' electrons in
UPd$_2$Al$_3$. {\it Nature} 410:340-343.

\bibitem{joe} Sarrao JL and Thompson JD (2007) Superconductivity in
cerium- and plutonium-based `115' materials. {\it J. Phys. Soc.
Jpn.} 76:051013.

\bibitem{broholm} Stock C, Broholm C, Hudis J, Kang HJ, Petrovic C
(2008) Spin resonance in the d-wave superconductor CeCoIn$_5$.
{\it Phys. Rev. Lett.} 100:087001.

\bibitem{super} Sidis Y, {\it et al.} (2004) Magnetic resonant
excitations in high-$T_c$ superconductors. {\it phys. stat. sol.
(b)} 241:1204-1210.

\bibitem{sup1} Hayden SM, Mook HA, Dai PC, Perring TG, Dogan F
(2004) The structure of the high-energy spin excitations in a
high-transition-temperature superconductor. {\it Nature}
429:531-534.

\bibitem{pham} Pham LD, Park T, Maquilon S, Thompson JD, Fisk Z
(2006) Reversible tuning of the heavy-fermion ground state in
CeCoIn$_5$. {\it Phys. Rev. Lett.} 97:056404.

\bibitem{kato} Kato M and Machida K (1988) Superconductivity and
spin-density waves - Application to Heavy-fermion materials. {\it
Phys. Rev. B} 37:1510-1519.

\bibitem{curro} Urbano RR, {\it et al.} (2007) Interacting
Antiferromagnetic Droplets in Quantum Critical CeCoIn$_5$. {\it
Phys. Rev. Lett.} 99:146402.

\bibitem{nicklas} Nicklas M, {\it et al.} (2007) Magnetic structure
of Cd-doped CeCoIn$_5$. {\it Phys. Rev. B} 76:052401.

\bibitem{bao} Bao W, {\it et al.} (2000) Incommensurate magnetic
structure of CeRhIn$_5$. {\it Phys. Rev B} 62:R14621-R14624.

\bibitem{bianchi} Bianchi A, Movshovich R, Capan C, Pagliuso PG,
Sarrao JL (2003) Possible Fulde-Ferrell-Larkin-Ovchinnikov
superconducting state in CeCoIn$_5$. {\it Phys. Rev. Lett.}
91:187004.

\bibitem{science} Kenzelmann M, {\it et al.} (2008) Coupled
superconducting and magnetic order in CeCoIn$_5$. {\it Science}
321:1652-1654.

\bibitem{yoshi} Tokiwa Y, {\it et al.} (2008) Anisotropic effect
of Cd and Hg doping on the Pauli limited superconductor
CeCoIn$_5$. {\it Phys. Rev. Lett.} 101:037001.

\bibitem{Cr1} Jaramillo R, {\it et al.} (2007) Microscopic and
macroscopic signatures of antiferromagnetic domain walls. {\it
Phys. Rev. Lett.} 98:117206.

\bibitem{Cr2} Kummamuru RK, Soh Y-Ah (2008) Electrical effects
of spin density wave quantization and magnetic domain walls in
chromium. {\it Nature} 452:859-863.

\bibitem{kas} Kasahara Y, {\it et al.} (2005) Anomalous
quasiparticle transport in the superconducting state of
CeCoIn$_5$. {\it Phys. Rev. B} 72:214515.

\bibitem{sunil} Nair S, {\it et al.} (2009) Hall effect and
magnetoresistance in the heavy fermion superconductor
CeCo(In$_{1-x}$Cd$_x$)$_5$. {\it J. Phys.: Conf. Series}
150:042133.

\bibitem{cur2} Curro NJ and Pines D (2007) Anisotropic spin
fluctuations in CeCoIn$_5$ {\it J. Phys. Chem. Solid}
68:2028-2030.

\bibitem{paschke} Paschke C, Speck C, Portisch G, von L\"ohneysen
H (1994) Magnetic-ordering and Kondo compensation in the ternary
heavy-fermion compound CeCu$_5$Au. {\it J. Low Temp. Phys.}
97:229-250.

\bibitem{geg} Gegenwart P, {\it et al.} (1998) Breakup of heavy
fermions on the brink of `phase A' in CeCu$_2$Si$_2$. {\it Phys.
Rev. Lett.} 81:1501-1504.

\bibitem{brandt} Brandt NB and Moshchalkov VV (1984) Concentrated
Kondo systems. {\it Adv. Phys.} 33:373-467.

\bibitem{dahm} Dahm T, {\it et al.} (2009) Strength of the
spin-fluctuation-mediated pairing interaction in a
high-temperature superconductor. {\it Nature Phys.} 5:217-221.
\end{thebibliography}
\end{document}